\UseRawInputEncoding
\documentclass[a4paper,10pt,oneside]{article}
\usepackage{icad2026,amsmath,epsfig,times,url,hyperref}

\title{REAL-TIME, EDM-INSPIRED SONIFICATION OF THE ACTIVITY OF A SUPERCOMPUTER}

\twoauthors{Marco Alunno} {Universidad EAFIT \\ Carrera 49 \#7sur-50 \\ Medellín, Colombia  \\ {\tt malunno@eafit.edu.co}}
{Paolo Bientinesi} {Ume\aa Universitet \\ Universitetstorget 4, 901 87 \\ Ume\aa, Sweden  \\ {\tt pauldj@cs.umu.se}}

\usepackage{xcolor}

\usepackage{booktabs}

\begin{document}
\ninept
\maketitle

\begin{sloppy}

\begin{abstract}
The project described in this paper explores the informative sonification of data received in real time from a supercomputer. These data capture the current activities in all the nodes of the computer, therefore, their sonification functions as a form of continuous monitoring of the nodes' behavior and, by extension, of the system as a whole. Because such monitoring is theoretically unending, the resulting sonification must be musically capable of conveying information through sound in a way that remains both intelligible and engaging over long durations. Rather than imposing a predefined musical style onto the data, we sought to identify one which the data themselves could plausibly support. From a small set of candidates, we selected EDM because it is a family of genres whose structural and temporal characteristics align well with continuous, data-driven processes and long-term listening. Through this style-based approach, this research builds on the long tradition of computer data sonification while uniquely combining three elements rarely addressed together: monitoring (rather than debugging) as the primary goal, real-time (rather than post-mortem) data interpretation, and generation of virtually infinite and stylistically coherent (rather than incongruous) music.    
\end{abstract}

\section{Introduction}
\label{sec:intro}

This project is driven by a conceptual metaphor: A supercomputer, with its thousands of cores relentlessly executing computational tasks and perfectly coordinated by a central management system, evokes the image of an immense mechanized orchestra under the control of a conductor's baton.
Further, a supercomputer is organized according to a hierarchy in which the individual cores are part of nodes, which, depending on their typology, are in turn grouped into partitions.  
Within the metaphor, if a supercomputer is an orchestra, the partitions are the orchestra’s instrumental families, and the nodes are the players whose activity is controlled by the management system (conductor).
Since the nodes' activity is summarized through a set of features expressed as numerical values, converting these values into sound would produce an acoustic image of the behavior of each node and, consequently, a radiograph of the computer as a whole. The goal of this work is to design such a conversion and to compose an informative sonification for the acoustic monitoring of a supercomputer that is in real time, uninterrupted, and never-ending.

Monitoring refers to the continuous, low-overhead observation of a running system aimed at understanding its state, performance, and long-term trends. Unlike debugging, monitoring does not involve the investigation, identification, or correction of abnormal behaviors. In our case, the monitored events are the activities and processes of a supercomputer, which 
are traditionally presented through data visualization techniques~\cite{MonitoringSystems}. 
Because monitoring may extend over long periods of time, offloading the representation of data from the visual to the auditory domain can be advantageous: as noted in prior works~\cite{walker2011theory, vickers2011process, brown1992color}, a sonification reduces the need for undivided visual attention and allows the listener to remain engaged without continuously focusing on a screen.
As target listeners, we picture supercomputer users and administrators, as well as ordinary passersby in spaces around the supercomputer facilities. 

The supercomputer we chose to monitor, Kebnekaise, is operated by the High-Performance Computing Center North (HPC2N) at Ume\aa\ University, Sweden. At the time of writing, Kebnekaise comprises 10
partitions, 95 nodes (for a total of 206 CPUs and 74 GPUs), whose activities are overseen by the Simple Linux Utility for Resource Management (Slurm) workload manager.
Slurm continuously monitors the state of the processors of each node, and schedules incoming requests to available resources (Sec.~\ref{sec:usage}). Additionally, Slurm has also the ability to collect various data, and to produce informative statistics about usage, conditions, and activities of the nodes. In our study, we focus on three metrics---namely, the number of running processes, the amount of physical memory used, and the amount of input/output activity. In detail, information about these metrics is periodically extracted from each node, gathered into three arrays in a local server, and then reorganized as three streams that are sent to the internet. 

The three streams are received by a different computer where a Python script reads and forwards them, via OSC, to the sonification code in SuperCollider\footnote{The sonification code in SuperCollider is available at \hyperlink{}{https://github.com/pupil72/kebne-sonification}}.
A central feature of the project, that enables users to gain an immediate sense of the levels of activity within Kebnekaise, is that the sonification refreshes in real time with the incoming data stream.   Moreover, the regular, periodic renewal of the sonification
leverages anticipation as a perceptual mechanism, helping to sustain the listener’s engagement during continuous monitoring~\cite{huron2006sweet}.

Sustained engagement is particularly important in this project, because the duration of the sonification is undefined, lasting at least until the supercomputer is restarted, which normally happens every few weeks or months.
However, engagement is not a direct consequence of the anticipation effect. For anticipation to be operative, it must allow the listener to form predictions---however correct or incorrect---about forthcoming sonic events. Consequently, for sound to be engaging, it must be organized in a way that encourages predictive listening. When the listener possesses prior knowledge of that organization, engagement may arise immediately; otherwise, the principles governing the organization may gradually emerge through extended exposure~\cite{hauer2015sonification}. Our search of suitable forms of organization ultimately led us to conceive sound in terms of musical style---that is, sound structured according to a set of rules, possibly known by most, that constrain and define its behavior. The preconditions of regular changes and indefinite length suggested us to use a repetitive, non-teleological musical style. Out of a shortlist of candidates, among which drone and ambient music, we soon identified EDM as the most appropriate, since it subsumes an umbrella of genres whose formal structure and temporal behavior are well suited to continuous, data-driven processes and extended listening. In fact, many real-world applications of EDM already support the practice of extended listening, as this style of music is primarily intended for use in dance-oriented settings that can last from several hours to an entire day or even longer.

In what follows, we first provide an overview of previous research on the sonification of supercomputing. We then briefly describe several aspects related to the mechanisms involved in handling and managing a supercomputer's resources. The next two sections discuss, respectively, the questions we addressed in order to advance the project and the solutions we ultimately adopted, together with the technical characteristics of the sonification. We conclude by outlining potential improvements and future extensions of the project.

\section{Literature review}
\label{sec:literature}

The development of sonification for computer and supercomputer processes has transitioned from the accidental detection of electromagnetic interference to the intentional design of musical frameworks for continuous system monitoring. In the mid-twentieth century, early engineers and programmers relied on unintended audio cues (e.g. tuning AM radios to pick up CPU interference or attaching speakers to index registers) to discern the state of running compilers~\cite{digiano1992program, vickers2003siren}. These primitive methods established the fundamental concept that computer activity possesses a distinct sound life that can be meaningfully interpreted by the human ear. However, while these early instances provided a form of monitoring, the field quickly moved toward structured "program auralization" focused primarily on debugging. Systems like CAITLIN were designed to map Pascal constructs to specific musical motifs, enabling programmers to hear if and when code deviated from expected paths~\cite{vickers2002when}. This approach, alongside the InfoSound toolkit and LogoMedia environment, prioritized code comprehension and the detection of rapid event sequences that were often visually overwhelming~\cite{sonnenwald1990infosound, digiano1992program}. Our research distinguishes itself from this tradition by shifting the focus away from error detection back to continuous monitoring as the primary goal, addressing systems that are theoretically unending rather than finite blocks of code.

As computer systems grew in complexity, the focus expanded to the challenges of monitoring supercomputers and parallel architectures. Research into the scalability of visual-aural representations demonstrated that while visual displays often become cluttered as the number of processors increases, sound remains a robust medium for portraying large volumes of parallel data~\cite{francioni1991debugging, francioni1992visual}. By mapping processor activity to sound, researchers identified "aural signatures"---unique patterns representing specific communication topologies---allowing operators to detect load imbalances and network congestion~\cite{jackson1992aural}. Despite these advancements, much of this foundational work relied on forensic data interpretation, often using trace files recorded during execution rather than live streams~\cite{francioni1991debugging, jackson1992aural}. Our project offers a unique contribution by focusing on real-time data interpretation, capturing the current activity of all the nodes in the supercomputer as it unfolds, rather than processing data post-mortem.

The requirement for long-term monitoring led to the exploration of real-time auditory displays and ambient soundscapes to provide peripheral awareness. Systems like Personal Webmelody and those developed for streaming server logs were designed to give administrators a background sense of system health~\cite{barra2002multimodal, hauer2015sonification}. The concept of Weakly Intrusive Ambient Soundscapes (WISP) further refined this by using nature-inspired models, such as the Peep Network Auralizer, which used recordings of rain or waterfalls to represent network conditions~\cite{kilander2002whisper, gilfix2000peep}. While these systems sought to be non-intrusive, they often turned music into an annoying "bleep bloop"~\cite{ballora2010preliminary}
or an aural object with no formal musical structure. Our research addresses this limitation by generating virtually infinite and stylistically coherent music inspired on EDM. This style-based approach leverages rhythmic and temporal characteristics that are inherently aligned with continuous data-driven processes and algorithmic generation, ensuring that sonification remains both intelligible and engaging over long durations~\cite{anderson2013gedmas, savery2018interactive}.

In recent years, the aesthetics of sonification have become a central concern, with advanced frameworks exploring glitch music and unique acoustic timbres to represent hidden computer processes such as idle modes and shutdowns~\cite{panariello2022sonification, frisk2021sonifying}. While these artistic realizations raise awareness of digital complexity, they are typically applied to discrete, finite events~\cite{panariello2022sonification}. Furthermore, while direct sonification frameworks can handle large multivariate datasets by synthesizing sound grains, they often prioritize pattern recognition over long-term musical engagement~\cite{krekovic2017towards}. By combining the goal of continuous monitoring with real-time interpretation and a commitment to a specific musical style, our work builds upon the long tradition of computer data sonification while uniquely addressing a gap in the creation of informative displays that are musically sophisticated enough for indefinitely long listening.

One aspect that this tradition of data sonification shows is the frequent adoption of parameter mapping strategies, where multivariate data dimensions are mapped to complex sonic attributes~\cite{Worrall2010BetterPerception}. Early validation of this approach, such as Yeung’s (1980) work on pattern recognition in analytical data, demonstrated that the human ear is highly capable of identifying data-driven regularities when mapped to independent properties of sound~\cite{GrondBerger2011ParameterMapping, Yeung1980PatternRecognition}. However, researchers have highlighted that the override of aesthetics to ensure structural clarity, a sonification strategy that Worral named 'soniculation'~\cite{Worrall2010BetterPerception}, clashes with data-driven music composition, in which the expression of cultural and musical knowledge is the primary goal.
Our project navigates this divide by adopting EDM as a functional yet culturally situated style, aiming for a "pragmatic information aesthetic" that balances analytical utility with long-term listener enjoyment~\cite{BarrassVickers2011SonificationDesign}.

The transition to real-time supercomputer monitoring aligns with the principles of "systemic sonification," which reframes the auditory display as a living system that responds dynamically to its environment~\cite{Seica2021SystemicSonificationAesthetics}. This perspective moves beyond the "pure meaning" utopia of functionalism to embrace the subjective, embodied nature of the listening experience~\cite{Seica2021SystemicSonificationAesthetics}. In fact, as Morabito suggests, by shifting the focus from information transmission to the aesthetic and symbolic form, sonification can foster a "community without communication," where shared practice and narratives become central. Through this ritualistic approach to sonic interaction design, the sonification becomes a performative act where the "signifier"--—in this case, the rhythmic structure of the EDM track—--is appreciated alongside the data it conveys~\cite{Morabito2022RitualisticApproach}. This approach encourages the development of "aural fluency," where the listener gradually internalizes the logic of the soundscape, allowing the supercomputer’s hidden activity to become a part of the user’s lived environment through micro-rituals of presence~\cite{Vickers2011ProcessMonitoring, Morabito2022RitualisticApproach}.

Finally, the choice of an auditory medium addresses the critical challenge of cognitive load in complex work domains. As noted by Francioni et al., the human ability to process audio information in a passive manner allows for situational awareness without requiring constant active monitoring~\cite{francioni1991debugging}. In environments where visual displays are often overtaxed, a well-designed auditory backdrop can provide a "calm technology" that informs without distracting~\cite{Vickers2011ProcessMonitoring}. By leveraging the passive processing capabilities of the brain, our sonification enables administrators to detect subtle changes or anomalies in the supercomputer's high-dimensional stream while maintaining focus on other primary tasks, effectively turning the data into an informative and "decorative" background soundscape~\cite{hauer2015sonification}.

\section{Usage of a supercomputer}
\label{sec:usage}
There exists no formal definition to tell apart a computer from a supercomputer. Loosely speaking, the distinction separates those computers that are portable or fit under a desk from those that need a dedicated facility, and likewise, those that are used by one or just a few concurrent users from those that accommodate thousands of concurrent users and jobs. Furthermore, access to supercomputers is normally carefully regulated, and individual users can only make use of a (small) portion of the total amount of available resources. From the user's perspective, once access to the supercomputer has been granted, the workflow looks as follows: 1) A description of the task that one wants to execute is prepared; such description includes the name of the program that is to be executed, the parameters and the environment for the program, and most notably, a list of resources that the program should make use of (e.g., number of cores, number of nodes, type of nodes, amount of memory, amount of time, etc.). 2) The description is submitted to a resource manager (e.g., Slurm) and becomes a standing request. This request is assigned a certain priority, and sits in a queue alongside thousands of other requests that were submitted by the same and other users. 3) After a (hopefully short) wait, the request is eventually fulfilled, i.e., the program is scheduled to execute on the requested resources. 4) Once the program completes, its output is made available to the user (e.g., as a file or by email). 

For this workflow to function properly, it is essential that the resource manager constantly maintains one or more priority queues, and most importantly, keeps track---"monitors"---the state of each node of the supercomputer (e.g., "idle", "allocated", "draining", "down"). Whenever enough resources are available to satisfy the demands of a high-priority request, the program is scheduled, the request is removed from the queue, and the state of the nodes is updated. The overall mechanism is extremely complex due to the competing needs to allocate as many concurrent jobs as possible---thus avoiding the computer to sit idle---and to ensure a fair allocation of resources---so that requests that need large amounts of resources are not indefinitely overtaken by smaller allocations. 

As mentioned in the Introduction, in addition to its current state, each node has the possibility to provide information and statistics for many parameters. In the next two sections, we describe the sonification of three streams of data, namely:  
the number of processes currently running on the node (\textbackslash procs), the percentage of physical memory in use (\textbackslash memusage), and the amount of data transmitted (outgoing traffic) over the InfiniBand high-speed interconnect (\textbackslash IB-tx).

\section{Toward a sonification of Kebnekaise}
\label{sec: Toward the sonification of a supercomputer}

The sonification design presented in this article is the last of a long series of continuous refinements. We propose here an account of some challenges we encountered while sonifying Kebnekaise and we highlight the options we considered, aware that other possible solutions could have been found. The final outcome of our research is revealed in the next section.

%kinds of data
To begin with, the three types of data introduced at the end of the previous section are a selection of many alternatives we could have considered, for example the CPU frequencies, the power consumption or the number of idle and allocated nodes, just to name a few. The reasons why we selected those three specific types of data are to be found in the compromise between the need for basic, i.e., easily interpretable information, and the effort to gather and summarize it for each node. 

Moreover, given the hierarchical architecture of a supercomputer, we questioned the appropriate level of granularity at which to sonify Kebnekaise. 
In general, a supercomputer is organized into a handful of partitions, each consisting of tens or hundreds of nodes; each node contains 1 or more CPUs, and between 0 to a few GPUs, for a grand total of (many) thousands of cores.
To sonify at the core level would entail mapping the three features mentioned above onto a huge number of individual elements. At the node level, by contrast, the scale would be significantly reduced. At the partition level, finally, the number of elements would shrink to just a small amount.

%pondering granularity
We discarded the core-level granularity as excessively detailed. Although this volume of data might lend itself to interesting musical works like the grainy, stochastic compositions of Iannis Xenakis or Barry Truax, it would likely obscure any informative intent behind a dense curtain of sounds. Instead, the orchestral metaphor outlined in the Introduction appeared well suited to a node-by-node sonification, with each node within a partition corresponding to an individual member of the instrumental family associated with that partition (e.g., a flute within the winds section). The partition level, although seemingly too coarse for nuanced musical articulation, would be particularly appropriate for informative applications, given the smaller number of perceptually distinguishable elements. Nevertheless, it could also prove musically viable if enriched with sufficiently complex sound objects to compensate for the reduced data resolution. This is the solution we finally adopted.

%temporal granularity
Another kind of granularity concerns the refreshing time of the sonification. In theory, since data are sonified in real time, the faster they are gathered by Slurm the better. In reality, though, since monitoring should not hinder the supercomputer's performance, there is a practical limit imposed by the resources that the administrator of Kebnekaise judged reasonable to allot for our project. Eventually we agreed on receiving a new batch of data every 15 seconds. Each new batch is averaged with the previous one (to limit the impact of sudden changes in activity levels) and is sonified in real time.

%parameter mapping
Since we considered parameter-mapping sonification to be the most intuitive approach for our aims, a challenge was to decide what musical parameters to map that would be at the same time musically expressive, perceptively independent, and cognitively interpretable. We found that certain decisions would condition, for example, where to present our sonification and how it would sound. Thus, panning, which is otherwise a perceptively meaningful and measurable sound parameter, becomes musically inexpressive if listened in places where loudspeakers are not located on opposite sides, for example along the university's corridors. Likewise, a parameter like pitch, typically used for mapping because of our inborn sensitivity to frequency changes, may lack a cognitive attribution if applied to unpitched percussion.

\section{EDM as sonification of Kebnekaise}
\label{sec: EDM as sonification of a supercomputer}

We describe here our approach to the sonification of the data yielded by Kebnekaise and the strategy we applied to convert those data into EDM.

First of all, the shift from the orchestral vision to an EDM-based context required a reformulation of the original metaphor. Rather than sonifying individual nodes within each partition as separate instruments of an orchestral family, we decided to sonify each partition as a distinct layer of an EDM track \footnote{We use the word 'track' as a synonym of 'piece of music' or 'song'.}.  Using a parameter mapping technique, each layer would then have some parameters whose values are modulated by the values of all the nodes within a given partition. This means that when new data about running processes, physical memory, and I/O activity for each node becomes available, it is gathered, processed and sonified in real time.

At this point, it is useful to recall that an EDM track is consistently structured in a 4/4 meter, with musical periods spanning four bars and events aligned to even subdivisions of each bar, the smallest frequently being the 16th. Since Slurm delivers a new batch of data to be sonified approximately every 15 seconds, fitting the four beats of a bar into this interval yields a metronome of 128 BPM, which conveniently lies between a fast house and a slow techno tempi. Furthermore, in a typical studio production, the layers of an EDM track are usually organized according to the specific frequency band they occupy or the musical function they fulfill~\cite{Butler2006UnlockingGroove, snoman2019dance, Smith2021EDM}. Therefore, an EDM track has a rhythmic layer composed of several sub-layers spanning the full frequency spectrum (e.g. kick drum, snare drum, clap, etc.), as well as additional ones, including vocal, transitional (i.e., buildups and breakdowns), low-, mid- and high-frequency layers~\cite{SolbergJensenius2016}. Finally, EDM tracks are generally highly repetitive and based on rhythmical patterns, therefore our sonification organizes sound onsets for each layer in patterns that are repeated until a new batch is delivered. 

\begin{table}[t!]
  \begin{center}
    \renewcommand{\arraystretch}{1.2}
    \begin{tabular}{@{}ll c l@{}}
      \toprule
      \multicolumn{2}{l}{Partition name} & 
      \# nodes & 
      \multicolumn{1}{l}{Matching layer} \\
      \midrule
    &{\tt cpu\_largemem}       & 8  & Bass  \\     
    % 4 CPUs x 18 cores -> 32 CPUs = 8 * 72 cores
    &{\tt cpu\_sky}            & 48 & Female voice  \\  % 2 CPUs x 14 cores -> 96 CPUs = 28 * 48 cores 
    &{\tt cpu\_zen3}           & 1  & Male voice \\ 
    % 2 CPUs x 64 = 128 cores    
    &{\tt cpu\_zen4}           & 8  & Chords       \\   
    % 2 CPUs x 128 -> 16 CPUs = 256 cores      
    &{\tt \em gpu+cpu\_sky}  & 10 & Kick drum  \\
    % 2 CPUs x 14 cores -> 20 CPUs = 20 x 14 cores
    % 2 GPUs -> 20 GPUs
    &{\tt \em gpu+cpu\_zen3} & 3 & Sub bass \\
    %%% 2 CPUs x 24 -> 6 CPUs = 3 x 48 cores
    %%% 2 GPUs -> 6 GPUs
    &{\tt \em gpu+cpu\_zen4} & 13 & Snare    \\     
    %%% 2 CPUs x 64 cores -> 26 CPUs = 26 x 64 cores
    %%% 2 GPUs -> 26 GPUs
    &{\tt \em gpu\_2xh100+cpu\_zen4}& 1 & Shaker \\
    %%% 2 CPUs x 64 cores + 2 GPUs
    &{\tt \em gpu\_6xl40s+cpu\_zen4}& 2 & Hi-hats \\   
    % 2 CPUs x 64 -> 4 CPUs = 256 cores 
    % 6 GPUs -> 12 GPUs
    &{\tt \em gpu\_8xa40+cpu\_zen4} & 1 & Clap   \\ 
    % 2 CPUs x 64 cores + 8 GPUs
    \bottomrule
    \end{tabular}
    \caption{Kebnekaise: partitions, number of nodes and associated instrumental layers. In italics, partitions containing GPUs.
    %CPUs: 16, 96, 26, 32, 20, 6, 2, 4, 2, 2 = 206
    %GPUs: 26, 20, 6, 12, 2, 8 = 74
    }
    \label{table:partitions}
    \end{center}
\end{table}

Layers and their associations are shown in Table~\ref{table:partitions}; from left to right: the names of the partitions (in italics, partitions containing GPUs), the number of nodes per partition, and the instrumental or vocal layer associated with each partition. As it can be observed, the important distinction in high-performance computing between central and graphic processing units (CPUs and GPUs) is preserved in the sonification by associating the six GPUs partitions to instruments belonging to the unpitched rhythmic layers of the track (the sub bass, although pitched, has a musical function and frequency spectrum closer to a kick drum). All sounds are samples, with the exception of the bass and chords layers, which are synthetically produced. 

Unlike \textbackslash memusage, which is already a percentage, we mapped \textbackslash procs and \textbackslash IB-tx to the range $[0,1]$ in order to facilitate their use in the modulation of musical parameters.
However, due to the possible range of their values, a straightforward linear scaling (from $[0, max]$ to $[0, 1]$) would not be satisfactory. In fact, while the upper bound ($max$) for \textbackslash IB-tx can in principle be determined (either by estimating or by measuring the maximum outgoing bandwidth of the InfiniBand interconnect), it is uncommon that scientific codes use I/O intensely for prolonged intervals of time. In practice, the observed values for \textbackslash IB-tx are often noticeably smaller than $max$. As a consequence, a linear scaling would result in an undesired compression of values towards 0.
The situation is even worse for \textbackslash procs, since no upper bound ($max$) can be established for this parameter. 

In order to map any incoming value $v$, regardless of its magnitude, to the interval $[0,1]$, we make use of a moving temporal window. This is a buffer in which we store
the last $n$ non-scaled values, including $v$, the value from the most recent batch. We also keep track of $max$, the largest value within the window, which is then used to rescale $v$ from $[0,max]$ to $[0,1]$. The same process is not needed for the minimum, since 0 is the smallest possible value for all three types of data. 
It could however be applied (thus creating a "zooming lens" over the data stream) should one want to magnify differences in an otherwise monotonous stream.

The choice of $n$, the size of the moving window, has a large influence on the value $max$, which, in turn, directly determines how incoming values are rescaled. In fact, depending on $max$, the same incoming value for \textbackslash procs (or \textbackslash IB-tx) might be rescaled to significantly different values.
Large values of $n$ (in the order of days, weeks or months), lead to a slow-moving $max$ and to a sonification which we refer to as "long-term". 
This kind of sonification should be preferred when aiming to represent Kebnekaise's activity levels without being concerned by how low or unvarying those levels might be.
By contrast, small values of $n$ (e.g., less than one hour), result in a frequently-changing $max$ and  smaller values than those observed over longer periods of time. 
This kind of sonification, which we refer to as "short-term", is useful for magnifying modest variations in activity levels. In fact, should the usage of the supercomputer's resources remain relatively stable for prolonged periods of time, the sonification would be extremely monotonous. 
The sound files included in this articles were generated with $n=8$, that is, a rather small temporal window.

Of the three types of data we collect, \textbackslash procs is unquestionably the most important, as it reflects whether or not there is any activity in the node. We interpret a scaled value for \textbackslash procs (scaled-procs) below 0.1 as 0 (no activity) and sonify it consequently with a special event: one hit in the instrument corresponding to the node followed by an echo effect and silence until a new batch of data comes in [\underline{\href{https://drive.google.com/file/d/1Cb7-kelmojVd7Yxe9rEjJK26Yc2IICWJ/view?usp=drive_link}{sound 1}}]. Otherwise, scaled values above 0.1 will control the density of the onsets of the instrument within the pattern length, although not their position, which is random (see below). From the listener's perspective, this means that the more there are active processes in the partition's nodes, the fuller is the rhythmical pattern and vice versa. Additionally, in order to make the variation of \textbackslash procs more evident, the pattern length is set to two bars (half a period) instead of one, thereby increasing the number of possible rhythmic positions.

Fig.~\ref{fig:pattern} shows the pattern for the kick drum. Blue dots on positions 0, 8, 16 and 24 indicate the basic pattern, that is, the pattern that will be played when scaled-procs = 0.1 [\underline{\href{https://drive.google.com/file/d/1eIorEU2VLjWf4_NKP8sVVGcP1JZiC-zx/view?usp=drive_link}{sound 2}}]. For scaled values between 0.1 and 1, in addition to the basic pattern, a proportional number of available onsets is randomly selected and recalculated at every new batch [\underline{\href{https://drive.google.com/file/d/1D2A2MpABaKguUavMdM_YMtqcvFedSH37/view?usp=drive_link}{sound 3}}]. %\procs = 0.56 in sound_4
When scaled-procs = 1, every dot in the pattern is a kick hit [\underline{\href{https://drive.google.com/file/d/13QFOddoPLGReexNwIr7yHeD4whL08ijm/view?usp=drive_link}{sound 4}}].

\begin{figure}[t]
\begin{minipage}[h]{1.0\columnwidth}
  \centering
  \centerline{\epsfig{figure=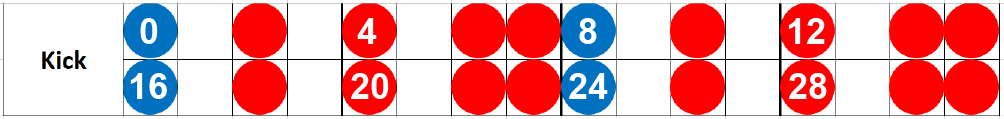,width=1\columnwidth}}
\end{minipage}
\caption{Kick drum pattern with blue dots indicating the basic pattern and red dots for other onset positions. Each square is a 16th subdivisions of the beat and thick vertical bars shows the start of a new beat (also marked by the numbered circles).
}
\label{fig:pattern}
\end{figure}

\textbackslash memusage controls the perceived pitch of the sound by affecting the playback speed of the samples. For those two partitions sonified with synthetic sounds, different parameters are used to achieve a perceptually similar effect. When  memusage = 0, samples are played at their original speed. Otherwise ($0 <$ \textbackslash memusage $\leq 1$), the playback speed increases proportionally across the pattern, starting with the first hit at normal speed and accelerating progressively with each subsequent hit. For the listener this corresponds to hearing a progression toward higher sounds when the nodes in the partition use more of their memory.

Finally, \textbackslash IB-tx is represented by the amount of reverberation applied to the sound, together with the amplitude of a delay added to reinforce this effect. When a scaled value for IB-tx (scaled-IB-tx) = 0, no reverb or delay is applied. Otherwise ($0 <$ scaled-IB-tx $\leq 1$), both the reverb width and the delay amplitude are proportionally larger for every successive hit in the pattern.

It is worth noting that the perception of each data stream (\textbackslash procs, \textbackslash memusage, and \textbackslash IB-tx) requires different listening strategies. \textbackslash IB-tx is immediately perceived at the first hit of a pattern, since the same amount of reverb and delay is applied uniformly to all hits, making the information effectively redundant over time. By contrast, \textbackslash memusage and \textbackslash procs can only be apprehended by listening to the entire pattern: the former, by grasping the pitch difference between the first and last hit, the latter by appreciating the onset density across the pattern.

Once the entire sound structure---with its multiple layers and their respective features---is set up, another issue remains to be addressed: how to present it to the listener. Sonifying all layers simultaneously may be musically appropriate, but it can be informatively problematic due to masking effects and an excess of acoustic information. In fact, if we receive three values per node every 15 seconds, this implies that every minute we should sonify 95 * 3 * 4 = 1140 values, that is, 19 values per second, a quantity that exceeds any user's analytical abilities to discriminate them.
This is why the current sonification is designed to facilitate the acoustic analysis of data: each layer moves to the foreground for two consecutive batches (approximately 30 seconds, or two musical periods) in a round-robin fashion, while the remaining layers are randomly kept in the background or silenced altogether. The layers are played in this order: kick, snare, hi-hats, clap, shaker, sub-bass, female voice, bass, chords, and male voice. Before a new cycle starts afresh, all layers play at the same time for two batches [\underline{\href{https://drive.google.com/file/d/1YErR7SriND76Fh7tuGgFLQHHQZGXudtv/view?usp=drive_link}{sound 5}}].

From its outset, this project has focused on creating an automatic music generation system with which the user is not expected to interact, rather than a sonification environment with programmable features and configuration options. For this reason, we have been reluctant to provide a graphical interface for direct sound control.
Nevertheless, in order to foreground the informative dimension of the sonification, we implemented a simple GUI that enables the listeners to switch from round-robin execution to a full display of all instruments. In the latter mode, one can thus select which layer(s) to monitor, thereby focusing on the specific partition(s) of interest (see Fig.~\ref{fig:GUI}).

\begin{figure}[t]
\begin{minipage}[h]{1.0\columnwidth}
  \centering
  \centerline{\epsfig{figure=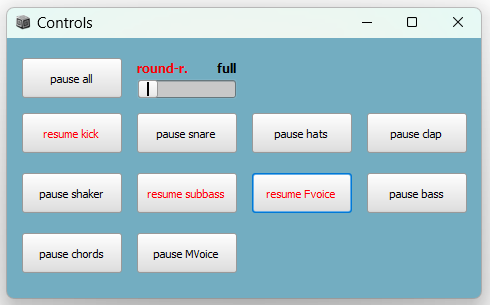,width=1\columnwidth}}
\end{minipage}
\caption{Sonification GUI, currently in round-robin mode with the kick, subbass and female voice paused.}
\label{fig:GUI}
\end{figure}

\section{Conclusions and Future works}
\label{sec:conclusions}
In this project, we have developed a system for the real-time, EDM-inspired sonification of the Kebnekaise supercomputer, aimed at providing continuous, informative monitoring. Our approach evolved from a conceptual "mechanized orchestra" metaphor into a stylistically coherent musical framework, where each supercomputer partition is represented as a distinct layer in an EDM track. This style was specifically selected because the repetitive, non-teleological structure and temporal characteristics of many genres it embraces conveniently align with data-driven processes and long-term listening. By mapping three key metrics—--running processes, physical memory usage, and transmitted data—--to corresponding sound features, we have created an auditory display that, we believe, remains both intelligible and engaging. Furthermore, to mitigate cognitive overload and information masking caused by the high volume of sonified data, we implemented a round-robin presentation strategy and a simple GUI that allows for the selective foregrounding of specific partitions.

This project is only part of a larger sonification effort. In fact, hitherto, we only tackled one half of the problem, namely, the informative side of the sonification. Our focus now shifts towards producing an artistic sonification that generates real-time, uninterrupted, never-ending EDM for dance purposes. By using the same data as in the informative sonification, we intend to interpret them prioritizing the emulation of a music style, rather than the perceptibility of the information they encode.

The sonification should also be accessible to the outer world instead of being confined to the local users of Kebnekaise. To this end, we plan to develop a web interface featuring a media player that allows users to switch between the informative and the artistic rendering of the data. Along this line, a further enhancement would be to enable users to listen to the sonification of their own program executions, rather than that of the entire system.

Lastly, in light of the fact that the Slurm workload manager is widely used in supercomputing facilities, we also aim to extend the project to other supercomputers. Indeed, by interfacing with Slurm’s accounting and monitoring tools, the same or similar metrics can be captured on different systems, thus enabling both broader dissemination and cross-facility comparison of activity. This transference to other supercomputers can be easily implemented only to a certain extent. In fact, our sonification is designed to sonify up to ten partitions, leaving open the question of how to sonify systems with larger architectures. 

\section{ACKNOWLEDGMENT}
\label{sec:ack}
This project would have not been possible without the technical contributions of \AA ke Sandgren. 
The authors wish to also thank Micka\"el Zehren for his constructive feedback.  
This work was supported by the eSSENCE Programme under the Swedish Government’s Strategic Research Initiative and by the Universidad EAFIT. 

\bibliographystyle{IEEEtran}
\bibliography{refs2026}

\end{sloppy}
\end{document}